\documentclass[aps,prl,preprint,showpacs,preprintnumbers,amsmath,amssymb,superscriptaddress]{revtex4}%

\usepackage{graphicx}%
\usepackage{dcolumn}
\usepackage{amsmath}

\makeatletter
\def\btt#1{\texttt{\@backslashchar#1}}%
\DeclareRobustCommand\bblash{\btt{\@backslashchar}}%
\makeatother

\begin{document}

\preprint{PREPRINT (\today)}

\title{Muon spin rotation study of the ternary noncentrosymmetric superconductors Li$_2$Pd$_x$Pt$_{3-x}$B }

\author{Petra S.~H\"afliger}
\affiliation{Physik-Institut der Universit\"{a}t Z\"{u}rich,
Winterthurerstrasse 190, CH-8057, Z\"urich, Switzerland}
\author{Rustem Khasanov}
\affiliation{Physik-Institut der Universit\"{a}t Z\"{u}rich,
Winterthurerstrasse 190, CH-8057, Z\"urich, Switzerland}
\author{Rolf Lortz}
\affiliation{D\'epartement de Physique de la Mati\`ere Condens\'ee, Universit\'e de Gen\`eve, 24 quai Ernest-Ansermet, CH-1211 Gen\`eve, Switzerland}
\author{Alexander Petrovi\'c}
\affiliation{D\'epartement de Physique de la Mati\`ere Condens\'ee, Universit\'e de Gen\`eve, 24 quai Ernest-Ansermet, CH-1211 Gen\`eve, Switzerland}

\author{Kazumasa Togano}
\affiliation{National Institute for Materials Science, 1-2-1 Sengen, Tsukuba 305-0047, Japan}

\author{Chris Baines}
\affiliation{Laboratory for Muon Spin Spectroscopy, Paul Scherrer Institute, Switzerland}

\author{Bij\"orn Graneli}
\affiliation{Physik-Institut der Universit\"{a}t Z\"{u}rich,
Winterthurerstrasse 190, CH-8057, Z\"urich, Switzerland}

\author{Hugo Keller}
\affiliation{Physik-Institut der Universit\"{a}t Z\"{u}rich,
Winterthurerstrasse 190, CH-8057, Z\"urich, Switzerland}

\begin{abstract}
We investigated the superconducting state of the noncentrosymmetric superconductors Li$_2$Pd$_x$Pt$_{3-x}$B with superconducting transition temperature $T_c$= 5.16(8) K ($x$=2.25), 3.56(8) K ($x=1.5$) and 2.60 K ($x=0$) by  means of muon-spin rotation ($\mu$SR) and specific heat experiments. The $\mu$SR relaxation rate $\sigma_{sc}$ was found to be constant at low temperatures for all the compounds. Data taken at different magnetic fields show that the magnetic penetration depth $\lambda$ is field-independent for Li$_2$Pd$_{2.25}$Pt$_{0.75}$B and Li$_2$Pt$_{3}$B. The electronic contribution to the specific heat measured in  Li$_2$Pd$_{1.5}$Pt$_{1.5}$B and Li$_2$Pt$_{3}$B increases exponentially at the lowest temperatures. These features suggest that the {\it whole family} of Li$_2$Pd$_x$Pt$_{3-x}$B are single-gap $s$-wave superconductors across the entire doping regime. 
\end{abstract}

\pacs{74.70.Dd, 74.62.Fj, 74.25.Ha, 83.80.Fg}

\maketitle


Superconductivity in systems without inversion symmetry has recently attracted considerable interest, see e.g. Refs.~\cite{hayashi,bauer}. The basic pairing states for a superconductor can be generally classified as spin-singlet and spin-triplet states. Cooper pairing in the spin-singlet channel is based on time-reversal symmetry (Anderson's theorem)~\cite{anderson1} whereas for spin-triplet pairing also  the existence of an inversion center in the crystal lattice is necessary~\cite{anderson2}. However, the lack of an inversion center in the crystal lattice induces an antisymmetric spin-orbit coupling which is not destructive to special spin-triplet states as has been shown by Frigeri et al.~\cite{frigeri} in the case of CePt$_3$Si for instance. 
\\
Lately the noncentrosymmetric superconductors Li$_2$Pd$_3$B and Li$_2$Pt$_3$B have been discovered~\cite{togano,takeya,badica1,badica2}. This system presents an ideal playing field to study superconductivity without inversion symmetry since there is no sign of magnetic order or strong electron correlation effects~\cite{togano} such as that found in CePt$_3$Si for instance.  A significant effort was put into identifying the superconducting state using various experimental methods. While measurements of the penetration depth~\cite{yuan} and nuclear magnetic resonance (NMR) experiments~\cite{nmr} suggest the existence of spin triplet states and nodes in the gap of Li$_2$Pt$_3$B, specific heat data indicate conventional behavior as expected for an isotropic single-gap $s$-wave superconductor in Li$_2$Pd$_3$B and Li$_2$Pt$_3$B  compounds~\cite{takeya}.  Furthermore, our previous muon spin rotation ($\mu$SR) and magnetization  experiments on Li$_2$Pd$_3$B show no sign of an unconventional superconducting pairing state~\cite{khasanov,igor}.
\\
In order to shed light on this controversy we have performed  a systematic study of  Li$_2$Pd$_x$Pt$_{3-x}$B  at different doping levels $x$ choosing  the bulk-sensitive transverse-field muon-spin rotation (TF-$\mu$SR) technique. Therewith we  investigated the temperature and field dependence of the $\mu$SR relaxation rate $\sigma$ in Li$_2$Pd$_x$Pt$_{3-x}$B with $x$=2.25, 1.5 and 0. In addition, we measured the low-temperature specific heat of Li$_2$Pd$_{1.5}$Pt$_{1.5}$B and Li$_2$Pt$_{3}$B. 


Li$_2$Pd$_x$Pt$_{3-x}$B was prepared by two step arc-melting~\cite{togano} for various doping concentrations. Details of the sample preparation can be found elsewhere~\cite{khasanov}. The field-cooled magnetization of all the compounds was measured using a superconducting quantum interference device (SQUID) magnetometer in fields ranging from 0.5 mT to 1.6 T at temperatures between 1.75 and 10 K. Specific heat was measured from 6 K down to 0.1 K. 
\\ \\
TF-$\mu$SR experiments were performed at the $\pi$M3 beam line at the Paul Scherrer Institute, Villigen, Switzerland. The Li$_2$Pd$_{x}$Pt$_{3-x}$B ($x$=2.25, 1.5 and 0) samples were field cooled from above $T_c$ down to 30 mK in fields of 0.1, 0.02 and 0.01 T.  We performed temperature scans at low fields for all the samples. Field-dependent data  up to 1.5 T were taken  for Li$_2$Pd$_{2.25}$Pt$_{0.75}$B  and Li$_2$Pt$_{3}$B where each point includes a set of data  at $T>T_c$ and at 1 K after  cooling in the field.
The $\mu$SR signal was recorded in the usual time-differential way by counting positrons from decaying muons as a function of time. The field distribution $P(B)$ is then obtained from the Fourier transform of the measured time spectra $P(t)$. $P(t)$ is fitted with a Gaussian line and an oscillatory term  for the signal originating from the sample as well as for the signal belonging to the sample holder (silver plate). The second moment of $P(B)$  is related to the $\mu$SR relaxation rate $\sigma$ which contains the superconducting component $\sigma_{sc}$ as well as a contribution from the nuclear moments $\sigma_n$,
$\sigma^2=\sigma_{sc}^2+\sigma_n$.

In extreme type-II superconductors such as Li$_2$Pd$_3$B (Ginzburg -Landau parameter $\kappa\approx 27$~\cite{khasanov}), the magnetic field distribution $P(B)$ is almost independent of the coherence length $\xi$~\cite{brandt}. The magnetic penetration depth $\lambda$ can then be calculated  from the known value of $\sigma_{sc}$ using Brandt's formula describing the field variation in an ideal triangular vortex lattice~\cite{brandt-2003}:

\begin{eqnarray}
\sigma_{sc} [\mu\mbox{s}] & = & 4.83\times 10^4(1-h) \nonumber \\
& & \left[1+1.21\left(1-\sqrt{h}\right)^3\right]\cdot \lambda^{-2}[\mbox{nm}]
\label{brandt}
\end{eqnarray}
with $h=H/H_{c2}$ with $H_{c2}$ the uppper critical field.
Details can be found elsewhere, see e.g. Ref.~\cite{khasanov}.

\begin{figure}[htb]
 \centering\includegraphics[width=0.88\textwidth]{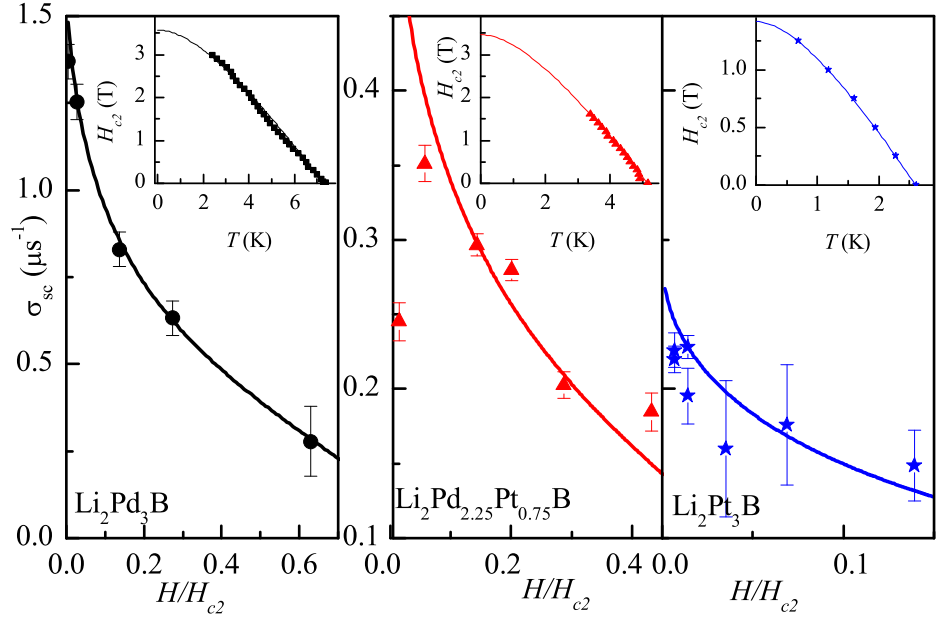}
 \caption{{Field dependence of the $\mu$SR depolarization rate $\sigma_{sc}(T=1\mbox{ K})$ in Li$_2$Pd$_{x}$Pt$_{3-x}$B~ for $x=3$ (left panel)~\cite{khasanov} as well as $x$=2.25 (middle panel) and 0 (right panel).  The solid lines correspond to the results of a least squares fitting procedure based on Eq.~\ref{brandt} as described in the text. In the insets the upper critical field $H_{c2}$ vs. temperature $T$ is shown with the solid lines corresponding to a fit based on the WHH model.}
\label{fscan}}
\end{figure}

The magnetic field dependence of the second moment of $\sigma_{sc}(1\mbox{ K})$ for  Li$_2$Pd$_{2.25}$Pt$_{0.75}$B and Li$_2$Pt$_3$B are displayed in the middle and right panel of Fig.~\ref{fscan}, respectively. For the sake of completeness we also show the field dependence of  $\sigma_{sc}(0)$  for Li$_2$Pd$_3$B~\cite{khasanov} in the left panel. The data are taken at a fixed  field at a  temperature $T>T_c$,  the sample is then field-cooled to 1 K  in order to determine $\sigma_{sc}$ and $\sigma_n$ unambiguously for each field. The insets illustrate the corresponding temperature dependence of the upper critical field $H_{c2}$ as obtained from magnetization (Li$_2$Pd$_3$B and Li$_2$Pd$_{2.25}$Pt$_{0.75}$B ) and specific heat measurements (Li$_2$Pt$_3$B). Clearly $H_{c2}(T)$ can be satisfactorily fitted with the model based on the Werthamer-Helfand-Hohenberg (WHH) theory~\cite{whh} as demonstrated by the solid lines. The results for $H_{c2}(0)$ are listed in Table~\ref{table}.
$H_{c2}$ is then inserted into Eq.~\ref{brandt} in order to fit $\sigma_{sc}(H)$ to the experimental data with $\lambda$ as a fit parameter. The results are in excellent agreement with the experimental data (shown by solid lines) implying  that $\lambda$ is field-independent. The deviation at low fields ($h\leq 0.003$ in the case of Li$_2$Pd$_{2.25}$Pt$_{0.75}$B) is expected since  Eq.~\ref{brandt} is no longer valid at this region. In the case of Li$_2$Pt$_3$B the error bars are large due to the small value of the relaxation rate $\sigma_{sc}$. The resulting penetration depth $\lambda(1\mbox{ K})$ equals  413(8) nm and 608(10) nm for Li$_2$Pd$_{2.25}$Pt$_{0.75}$B and Li$_2$Pt$_3$B, respectively, while we obtained $\lambda(0)=252(2)$ nm for Li$_2$Pd$_3$B~\cite{khasanov}, see Table~\ref{table}. The value of $\lambda(0)$ for Li$_2$Pd$_3$B is in agreement with magnetization measurements performed by Badica et al.~\cite{badica2} whereas our results for $\lambda(0)$ for Li$_2$Pd$_{2.25}$Pt$_{0.75}$B and Li$_2$Pt$_3$B deviate from those in Ref.~\cite{badica2}. 

The  superconducting coherence length $\xi$ may be estimated from $H_{c2}$ as 
$\xi = \left[\Phi_0/\left(2\pi H_{c2}\right)\right]^{0.5}$. We obtain $\xi(0)= 9.5(2)$ nm (Li$_2$Pd$_3$B)~\cite{khasanov}, $\xi(1\mbox{ K})=9.9(2)$ nm (Li$_2$Pd$_{2.25}$Pt$_{0.75}$B) and 15.2(8) nm (Li$_2$Pt$_3$B). This results in a value for $\kappa(0)=\lambda(0)/\xi(0)$ of  27(1) for   Li$_2$Pd$_{3}$B~\cite{khasanov} with $\kappa(1\mbox{ K})=42(1)$ and 40(1) for  Li$_2$Pd$_{2.25}$Pt$_{0.75}$B  and Li$_2$Pt$_{3}$B, respectively (see Table~\ref{table}). Since $\lambda$ is constant at low temperatures as will be shown later, we may set $\xi(0)\approx \xi(1\mbox{ K})$ as well as $\lambda(0)\approx \lambda(1\mbox{ K})$. The compounds in the present work are thus extreme type-II superconductors; $\mu$SR is therefore an appropriate tool for penetration depth studies.

\begin{figure}[htb]
 \centering\includegraphics[width=0.8\textwidth]{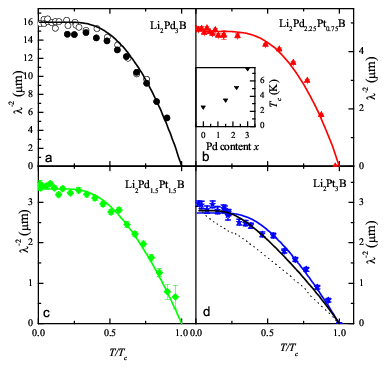}
 \caption{{Temperature dependence of $\lambda^{-2}$ in Li$_2$Pd$_x$Pt$_{3-x}$B for $x$=3 at 0.02 T (black open circles) and 0.5 T (black solid circles) (panel a), 2.25 (panel b) , 1.5 (panel c) and 0 (panel d) at 0.01 T.  The solid lines corresponds to a least-squares fit to the experimental data for a BCS single-gap $s$-wave superconductor based on Eq.~\ref{bcs} with the parameters given in the text. In panel d, the black solid and dotted line corresponds to model based on a two-gap scenario where one and both bands become superconducting, respectively, see text for explanation. The inset of panel b illustrates $T_c$ as a function of Pd doping $x$.}
\label{tscan}}
\end{figure}

We also measured the temperature dependence of the $\mu$SR signal in order to extract $\lambda^{-2}(T)$  using  Eq.~\ref{brandt}. The results are displayed in Fig.~\ref{tscan} for the entire doping series Li$_2$Pd$_x$Pt$_{3-x}$B ($x$=0, 1.5, 2.25 and 3~\cite{khasanov}). The second moment of $P(B)$ increases with $x$,  and $\lambda^{-2}(x)$ monotonically grows in turn with increasing Pd content.

We observe that $\lambda^{-2}(T)$ saturates at low temperatures and becomes $T$-independent for all the compounds. This indicates that the Fermi surface is fully gapped down to the lowest temperatures. Thus we 
applied a model for a BCS $s$-wave superconductor in order to fit the data (see e.g Ref.~\cite{kim})

\begin{eqnarray}
\frac{\lambda^{-2}(0)}{\lambda^{-2}(T)} = 1+2\cdot\left\langle\int_0^{\infty} d\epsilon \frac{\partial f}{\partial E}\right\rangle.
\label{bcs}
\end{eqnarray}

\begin{figure}[htb]
 \centering\includegraphics[width=0.8\textwidth]{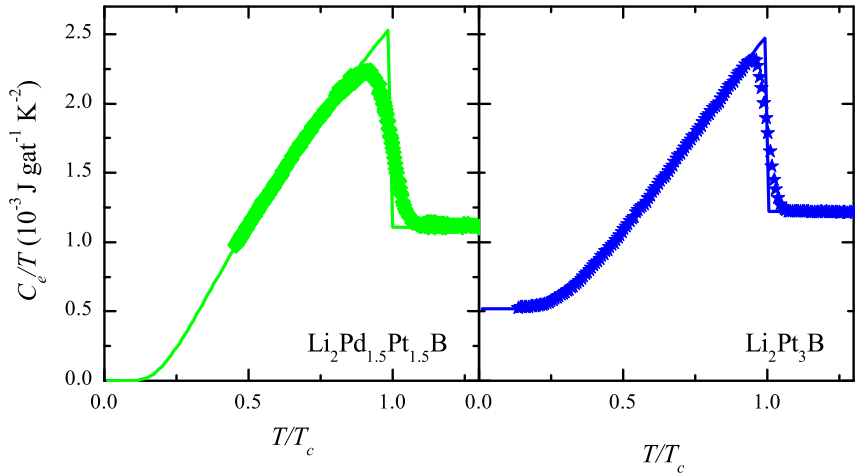}
 \caption{{Specific heat of Li$_{2}$Pd$_{1.5}$Pt$_{1.5}$B (left panel) and Li$_{2}$Pt$_{3}$B (right panel).} 
\label{spec-heat}}
\end{figure}

Here $\langle..\rangle$ represents an angular average over the Fermi surface, $f$ is the Fermi function and $E=\sqrt{\epsilon^2+|\Delta|^2}$ the total energy with $\epsilon$ the single-particle energy measured from the Fermi surface. Since we have assumed an isotropic gap, i.e. $s$-wave symmetry, the superconducting gap $\Delta$ is $k$-independent, but only depends on temperature $T$ such that  $\Delta(T)=\Delta_0\cdot \tanh(a\sqrt{T_c/T-1})$ with $a=1.74$~\cite{bonalde}. The resulting gap values at zero temperature $\Delta_0$ are listed in Table~\ref{table}. Over the whole temperature range ($T_c$ down to 30 mK) and in the whole doping range ($0\leq x\leq 3)$ $\lambda^{-2}$ is consistent with what is expected for an  isotropic BCS superconductor. 
However, the BCS fit to the Li$_2$Pt$_3$B data is not satisfactory. In Li$_2$Pt$_3$B a considerable spin-triplet contribution has previously been observed by Yuan et al.~\cite{yuan} and~\cite{nmr} thus we assumed a mixture of spin-triplet states associated with a strong antisymmetric spin-orbit coupling (ASOC) resulting in two bands as presented in Ref.~\cite{yuan}. Assuming the same set of parameters and a BCS-temperature dependence of the superconducting gaps (black dotted line in Fig.~\ref{tscan}d) $\lambda^{-2}(T)$ differs significantly from our results.   If we assume that the gap with nodes is smeared out due to inhomogeneity of the sample, the resulting model calculations are in agreement with our experimental data as illustrated by  the solid black line in Fig.~\ref{tscan}d. At this point we cannot exclude the existence of spin-triplet states.
\\ \\
In order to clarify matters, the superconducting state was also investigated by an independent study of the specific heat.   Fig.~\ref{spec-heat} illustrates the specific heat obtained for Li$_{2}$Pd$_{1.5}$Pt$_{1.5}$B (left panel)  and Li$_{2}$Pt$_{3}$B (right panel) in zero field.  The electronic contribution $C_e/T$ to the specific heat of Li$_2$Pt$_3$B, as obtained in a standard procedure by quenching superconductivity in a field of 4 T, shows an exponential increase at the lowest temperature. This is consistent with our $\mu$SR data and therefore implies that Li$_2$Pt$_3$B is a fully gapped superconductor. A fit using a standard single gap $s$-wave BCS model gives an excellent result (see Fig.~\ref{spec-heat}). We obtain a gap-to-$T_c$ ratio of $2\Delta_0/k_BT_c=4.0(2)$ which is slightly larger than the $\mu$SR value but still in reasonable agreement. A smaller value $2\Delta_0/k_BT_c=3.3(2)$ is derived for Li$_2$Pd$_{1.5}$Pt$_{1.5}$B (see Table~\ref{table}). 
A particularity of Li$_2$Pt$_3$B is a rather large residual Sommerfeld coefficient $\gamma_r$. A possible explanation is that within a two-gap scenario as presented by Yuan et al.~\cite{yuan} only one gap becomes superconducting, while the gap with nodes is smeared out due to inhomogeneity of the sample. However, we have measured another piece of the same large sample and we found a variation of $\gamma_r$ of approximately 30\%, but the superconducting contribution remains unchanged. This behavior is typical for macroscopic metallic inclusions which remain in the normal state. Details of the specific heat measurements will be published elsewhere.


Interestingly, results in the literature postulate several different superconducting ground states for the  
Li$_2$Pd$_{x}$Pt$_{3-x}$B family. In particular previous work on  Li$_2$Pt$_3$B  deviate substantially  from our conclusions. For instance, Yuan et al.~\cite{yuan} found a linear temperature dependence of the low-temperature penetration depth in Li$_2$Pt$_3$B by using a tunnel-diode based, self inductive technique. The results were interpreted on the basis of a parity-mixed state with two Fermi surfaces induced by  the lack of inversion symmetry coupled with a strong ASOC. The authors found that in Li$_2$Pt$_3$B one of the superconducting gaps changes sign due to the enhanced ASOC which  leads to line nodes in the superconducting gap and therefore explains the non-saturating penetration depth. None of our data - neither $\mu$SR nor specific heat - supports such a scenario. We believe that the difference in the experimental data  arises from the different experimental techniques. In Ref.~\cite{yuan} data are taken  by using a surface sensitive technique with the magnetic field only penetrating a distance $\lambda$ below the surface (200-600~nm, see Table~\ref{table}).  Muons, on the other hand, penetrate to much bigger
distances (fraction of millimeters) and therefore allow us to study the magnetic penetration depth in the bulk. The specific heat is also a bulk-sensitive technique. Bearing these in mind, the discrepancy between the results obtained in the present study and those reported in Ref.~\cite{yuan} can
be explained by the difference between the bulk and the surface properties of Li$_2$Pd$_x$Pt$_{3-x}$B in analogy with what was observed in conventional~\cite{zrb12} and unconventional~\cite{ybco} superconductors. We would also like to emphasize that our data are in excellent agreement with the specific heat data presented in Ref.~\cite{takeya}. Takeya et al. found values for $2\Delta_0/k_BT_c$ = 3.94(4) and 3.53 for Li$_2$Pd$_3$B and Li$_2$Pt$_3$B, respectively, which are in reasonable agreement with our results obtained from $\mu$SR measurements.

We will briefly  comment on the results of  Nishiyama et al.~\cite{nmr} obtained using the  nuclear magnetic resonance (NMR) technique which is also bulk-sensitive. They deduce the existence of triplet states in Li$_2$Pt$_3$B from the Knight shift. However, it is not clear why they use a $p$-wave model to fit the data for the spin relaxation rate.  Furthermore, they predict the behavior for an $s$-wave symmetry of the superconducting gap in Li$_2$Pt$_3$B  by using the value for the superconducting gap of Li$_2$Pd$_3$B which is more than three times bigger, see Table~\ref{table}. In addition to this the absence of a Hebel-Slichter peak in the spin relaxation does not necessarily imply that the sample under investigation has an anisotropic gap.  
\\ \\
In conclusion we have performed muon-spin rotation, specific heat and magnetization measurements on the ternary boride superconductor Li$_2$Pd$_{3-x}$Pt$_x$B. The main results are (i) the field dependence of the $\mu$SR relaxation rate is consistent with Brandt's formula, (ii) the upper critical field $H_{c2}$ obeys the WHH model, (iii) the temperature dependence of $\lambda^{-2}$ and (iv) the specific heat data are consistent with what is expected for a $s$-wave BCS superconductor. The temperature dependence of $\lambda^{-2}$ of Li$_2$Pt$_3$B can also be explained by  the existence of two bands induced by  an ASOC where only the one without nodes contributes to superconductivity due to inhomogeneity of the sample. But specific heat data indicate the existence of metallic inclusions. All our measurements and analysis strongly point to a single isotropic superconducting gap. No signature of an exotic superconducting order parameter was found.  Thus we suggest that the {\it whole family}  of Li$_2$Pd$_{x}$Pt$_{3-x}$B is a standard BCS superconductor with a single, isotropic gap.


\begin{table*}
\caption[~]{\label{table} Summary of the normal state and superconducting parameters for Li$_2$Pd$_{x}$Pt$_{3-x}$B
obtained in the present study. The meaning of the parameters is
 explained in the text.}
\begin{tabular}{lcc|cccc|cccccccccc} \hline\hline

&\multicolumn{2}{c}{}&\multicolumn{4}{c}{$\mu$SR}&\multicolumn{4}{c}{Specific heat}\\
Sample&$T_c$&$\mu_0H_{c2}$&
       $\lambda(0)$&$\kappa(0)$&$\Delta_0$&$2\Delta_0/k_BT_c$&
       $\gamma$&$\Delta C/\gamma T_c$&$\Delta_0$&$2\Delta_0/k_BT_c$&&\\
       &[K]&[T]&[nm]&&[meV]&&[mJ/gat K$^2$]&&[meV]&&&\\
 \hline
 Li$_2$Pd$_{3}$B                              &7.66(5)$^a$\footnotetext[1]{as taken from Ref.~\cite{khasanov}}& 3.66(8)$^a$&
                                              252(2)$^a$&27(1)$^a$&1.53(4)&4.4(2)&
                                              1.50(2)$^b$\footnotetext[2]{as taken from Ref.~\cite{takeya}} &2.0(3)$^b$&1.30(4)$^b$&3.94(4)$^b$&\\
 Li$_2$Pd$_{2.25}$Pt$_{0.75}$B                &5.16(8)&3.35(9)&
                                              413(8)&42(1)&1.03(4)&4.4(3)&
                                              --&--&--&--&\\
 Li$_2$Pd$_{1.5}$Pt$_{1.5}$B                  &3.56(8)&2.27(8)&
                                              --&--&0.59(2)&3.7(2)&
                                              1.10(2)&1.28(3)&0.37(2)&3.3(2)&\\
 Li$_2$Pt$_{3}$B                              &2.60(5)&1.42(6)&
                                              608(10)&40(1)&0.41(2)&3.4(3)&
                                              1.66(2)&1.75(2)&0.45(2)&4.0(2)&\\
 \hline \hline
\end{tabular}

\end{table*}

This work was partly performed at the Swiss Muon Source (S$\mu$S), Paul Scherrer Institute (PSI, Switzerland). We are grateful to N.~Hayashi and M.~Sigrist for helpful discussions as well as to H.~Takeya for providing us the samples Li$_2$Pd$_x$Pt$_{3-x}$, $x=$3, 2.25, 1.5 and 0. This work was supported by the Swiss National Science Foundation and the K.~Alex M\"uller Foundation.


\begin{thebibliography}{17}
\expandafter\ifx\csname natexlab\endcsname\relax\def\natexlab#1{#1}\fi
\expandafter\ifx\csname bibnamefont\endcsname\relax
  \def\bibnamefont#1{#1}\fi
\expandafter\ifx\csname bibfnamefont\endcsname\relax
  \def\bibfnamefont#1{#1}\fi
\expandafter\ifx\csname citenamefont\endcsname\relax
  \def\citenamefont#1{#1}\fi
\expandafter\ifx\csname url\endcsname\relax
  \def\url#1{\texttt{#1}}\fi
\expandafter\ifx\csname urlprefix\endcsname\relax\def\urlprefix{URL }\fi
\providecommand{\bibinfo}[2]{#2}
\providecommand{\eprint}[2][]{\url{#2}}

\bibitem[{\citenamefont{Hayashi et~al.}(2006)\citenamefont{Hayashi,
  Wakabayashi, Frigeri, and Sigrist}}]{hayashi}
\bibinfo{author}{\bibfnamefont{N.}~\bibnamefont{Hayashi}},
  \bibinfo{author}{\bibfnamefont{K.}~\bibnamefont{Wakabayashi}},
  \bibinfo{author}{\bibfnamefont{P.~A.} \bibnamefont{Frigeri}},
  \bibnamefont{and} \bibinfo{author}{\bibfnamefont{M.}~\bibnamefont{Sigrist}},
%
%
%
  \bibinfo{journal}{Phys. Rev. B} \textbf{\bibinfo{volume}{73}},
  \bibinfo{eid}{024504} 
  (\bibinfo{year}{2006}).

\bibitem[{\citenamefont{et~al.}(2004{\natexlab{a}})}]{bauer}
\bibinfo{author}{\bibfnamefont{E.~Bauer} \bibnamefont{et~al.}},
  \bibinfo{journal}{Phys. Rev. Lett.} \textbf{\bibinfo{volume}{92}},
  \bibinfo{eid}{027003} 
  (\bibinfo{year}{2004}).

\bibitem[{\citenamefont{Anderson}(1959)}]{anderson1}
\bibinfo{author}{\bibfnamefont{P.}~\bibnamefont{Anderson}},
  \bibinfo{journal}{J. Phys. Chem. Solids} \textbf{\bibinfo{volume}{11}},
  \bibinfo{pages}{26} (\bibinfo{year}{1959}).

\bibitem[{\citenamefont{Anderson}(1984)}]{anderson2}
\bibinfo{author}{\bibfnamefont{P.~W.} \bibnamefont{Anderson}},
  \bibinfo{journal}{Phys. Rev. B} \textbf{\bibinfo{volume}{30}},
  \bibinfo{pages}{4000} (\bibinfo{year}{1984}).


\bibitem[{\citenamefont{P.~Frigeri and Sigrist}(2004)}]{frigeri}
\bibinfo{author}{\bibfnamefont{P.~A.} \bibnamefont{Frigeri},
  \bibfnamefont{D.F.~Agterberg}}, 
  \bibinfo{author}{\bibfnamefont{A.} \bibnamefont{Koga}}
  \bibnamefont{and}
  \bibinfo{author}{\bibfnamefont{M.}~\bibnamefont{Sigrist}},
  \bibinfo{journal}{Phys. Rev. Lett.} \textbf{\bibinfo{volume}{92}},
  \bibinfo{pages}{097001} (\bibinfo{year}{2004}).




\bibitem[{\citenamefont{et~al.}(2004{\natexlab{c}})}]{togano}
\bibinfo{author}{\bibfnamefont{K.~Togano} \bibnamefont{et~al.}},
  \bibinfo{journal}{Phys. Rev. Lett.} \textbf{\bibinfo{volume}{93}},
  \bibinfo{eid}{247004} 
  (\bibinfo{year}{2004}).

\bibitem[{\citenamefont{et~al.}(2005{\natexlab{a}})}]{takeya}
\bibinfo{author}{\bibfnamefont{H.~Takeya} \bibnamefont{et~al.}},
  \bibinfo{journal}{Phys. Rev. B} \textbf{\bibinfo{volume}{72}},
  \bibinfo{eid}{104506} 
  (\bibinfo{year}{2005}).

\bibitem[{\citenamefont{et~al.}(2004{\natexlab{d}})}]{badica1}
\bibinfo{author}{\bibfnamefont{P.~Badica} \bibnamefont{et~al.}},
  \bibinfo{journal}{Appl. Phys. Lett.} \textbf{\bibinfo{volume}{85}},
  \bibinfo{pages}{4433} (\bibinfo{year}{2004}).

\bibitem[{\citenamefont{et~al.}(2005{\natexlab{b}})}]{badica2}
\bibinfo{author}{\bibfnamefont{P.~Badica} \bibnamefont{et~al.}},
  \bibinfo{journal}{J. Phys. Soc. Jpn.} \textbf{\bibinfo{volume}{74}},
  \bibinfo{pages}{1014} (\bibinfo{year}{2005}).

\bibitem[{\citenamefont{et~al.}(2006{\natexlab{b}})}]{yuan}
\bibinfo{author}{\bibfnamefont{H.~Q.~Yuan} \bibnamefont{et~al.}},
  \bibinfo{journal}{Phys. Rev. Lett.} \textbf{\bibinfo{volume}{97}},
  \bibinfo{eid}{017006} 
  (\bibinfo{year}{2006}).



\bibitem[{\citenamefont{Nishiyama et~al.}(2007)\citenamefont{Nishiyama,
  Y.Inada, and G.~Q.~Zheng}}]{nmr}
\bibinfo{author}{\bibfnamefont{M.}~\bibnamefont{Nishiyama}},
  \bibinfo{author}{\bibnamefont{Y.Inada}}, \bibnamefont{and}
  \bibinfo{author}{\bibfnamefont{G.}~\bibnamefont{Q.~Zheng}},
%
%
  \bibinfo{journal}{Phys. Rev. Lett.} \textbf{\bibinfo{volume}{98}},
  \bibinfo{pages}{047002} (\bibinfo{year}{2007}).

\bibitem[{\citenamefont{et~al.}(2006{\natexlab{c}})}]{khasanov}
\bibinfo{author}{\bibfnamefont{R.~Khasanov} \bibnamefont{et~al.}},
  \bibinfo{journal}{Phys. Rev. B} \textbf{\bibinfo{volume}{73}},
  \bibinfo{eid}{214528} 
  (\bibinfo{year}{2006}).

\bibitem[{\citenamefont{et~al.}(2006{\natexlab{c}})}]{igor}
\bibinfo{author}{\bibfnamefont{I.~L.~Landau} \bibnamefont{et~al.}},
  \bibinfo{journal}{Physica C} \textbf{\bibinfo{volume}{451}},
  \bibinfo{eid}{132} 
  (\bibinfo{year}{2007}).

\bibitem[{\citenamefont{Brandt}(1988)}]{brandt}
\bibinfo{author}{\bibfnamefont{E.~H.} \bibnamefont{Brandt}},
  \bibinfo{journal}{Phys. Rev. B} \textbf{\bibinfo{volume}{37}},
  \bibinfo{pages}{2349} (\bibinfo{year}{1988}).

\bibitem[{\citenamefont{Brandt}(2003)}]{brandt-2003}
\bibinfo{author}{\bibfnamefont{E.~H.} \bibnamefont{Brandt}},
  \bibinfo{journal}{Phys. Rev. B} \textbf{\bibinfo{volume}{68}},
  \bibinfo{pages}{054506} (\bibinfo{year}{2003}).

\bibitem[{\citenamefont{Werthamer}(1966)}]{whh}
\bibinfo{author}{\bibfnamefont{N.~R.} \bibnamefont{Werthamer et~al.}},
  \bibinfo{journal}{Phys. Rev.} \textbf{\bibinfo{volume}{147}},
  \bibinfo{pages}{295} (\bibinfo{year}{1966}).

\bibitem[{\citenamefont{Kim}(2002)}]{kim}
\bibinfo{author}{\bibfnamefont{M.-S.} \bibnamefont{Kim et al.}},
  \bibinfo{journal}{Phys. Rev. B} \textbf{\bibinfo{volume}{66}},
  \bibinfo{pages}{064511} (\bibinfo{year}{2002}).

\bibitem[{\citenamefont{Bonalde}(2000)}]{bonalde}
\bibinfo{author}{\bibfnamefont{I.} \bibnamefont{Bonalde et al.}},
  \bibinfo{journal}{Phys. Rev. Lett.} \textbf{\bibinfo{volume}{85}},
  \bibinfo{pages}{4775} (\bibinfo{year}{2000}).

\bibitem[{\citenamefont{KhasanovZrB12}(2005)}]{zrb12}
\bibinfo{author}{\bibfnamefont{R.} \bibnamefont{Khasanov et al.}},
  \bibinfo{journal}{Phys. Rev. B} \textbf{\bibinfo{volume}{72}},
  \bibinfo{pages}{224509} (\bibinfo{year}{2005}).

\bibitem[{\citenamefont{Khasanovybco}(2005)}]{ybco}
\bibinfo{author}{\bibfnamefont{R.} \bibnamefont{Khasanov et al.}},
  \bibinfo{journal}{cond-mat.supr.con},
  \bibinfo{pages}{0705.0577v1} (\bibinfo{year}{2007}).

\end{thebibliography}

\end{document}